\documentclass[aps,amsfonts]{ar2e}
\input epsf
\usepackage{graphics,epsfig}
\begin{document}
\def\rem{$\clubsuit$}
\def\MP{M_P}
\def\MPd#1{M_{P,#1}}
\newcommand{\etal}{{et~al.}}
\def\citeasnoun#1{\citet{#1}}
\newcommand\be{\begin{equation}}
\newcommand\ee{\end{equation}}
\def\bea{\begin{eqnarray}}
\def\eea{\end{eqnarray}}
\def\crt{\nonumber \\}
\def\eqn#1#2{\be \label{eq:#1} #2  \ee}
\def\CPbar{\hbox{{\rmCP}\hskip-1.80em{/}}}
\def\ctn#1{{{\cite{#1}}}}
\def\reffig#1{Figure \ref{fig:#1}}
\def\rfn#1{Eq. {({\ref{eq:#1}})}}
\def\rfs#1{Sec.~{\ref{sec:#1}}}
\def\rfss#1{Sec.~{\ref{ss:#1}}}
\def\rfsss#1{Sec.~{\ref{sss:#1}}}
\def\lref#1#2{{\bibitem{#1}#2}}
%
\def\np#1#2#3{Nucl. Phys. {\bf B#1} (#2) #3}
\def\pl#1#2#3{Phys. Lett. {\bf #1B} (#2) #3}
\def\prl#1#2#3{Phys. Rev. Lett. {\bf #1} (#2) #3}
\def\pr#1#2#3{Phys. Rev. {\bf #1} (#2) #3}
\def\aph#1#2#3{Ann. Phys. {\bf #1} (#2) #3}
\def\prep#1#2#3{Phys. Rep. {\bf #1} (#2) #3}
\def\rmp#1#2#3{Rev. Mod. Phys. {\bf #1}}
\def\cmp#1#2#3{Comm. Math. Phys. {\bf #1} (#2) #3}
\def\mpl#1#2#3{Mod. Phys. Lett. {\bf #1} (#2) #3}
\def\ptp#1#2#3{Prog. Theor. Phys. {\bf #1} (#2) #3}
\def\jhep#1#2#3{JHEP {\bf#1}(#2) #3}
\def\jmp#1#2#3{J. Math Phys. {\bf #1} (#2) #3}
\def\cqg#1#2#3{Class.~Quantum Grav. {\bf #1} (#2) #3}
\def\ijmp#1#2#3{Int.~J.~Mod.~Phys. {\bf #1} (#2) #3}
\def\atmp#1#2#3{Adv.~Theor.~Math.~Phys.{\bf #1} (#2) #3}
\def\ap#1#2#3{Ann.~Phys. {\bf #1} (#2) #3}
%
%
%
%
\def\IB{\relax\hbox{$\inbar\kern-.3em{\rm B}$}}
\def\IC{\relax\hbox{$\inbar\kern-.3em{\rm C}$}}
\def\ID{\relax\hbox{$\inbar\kern-.3em{\rm D}$}}
\def\IE{\relax\hbox{$\inbar\kern-.3em{\rm E}$}}
\def\IF{\relax\hbox{$\inbar\kern-.3em{\rm F}$}}
\def\IG{\relax\hbox{$\inbar\kern-.3em{\rm G}$}}
\def\IGa{\relax\hbox{${\rm I}\kern-.18em\Gamma$}}
\def\IH{\relax{\rm I\kern-.18em H}}
\def\IK{\relax{\rm I\kern-.18em K}}
\def\IL{\relax{\rm I\kern-.18em L}}
\def\IP{\relax{\rm I\kern-.18em P}}
\def\IR{\relax{\rm I\kern-.18em R}}
\def\IT{{\bf T}}
\def\IZ{\relax\ifmmode\mathchoice{\hbox{\cmss Z\kern-.4em Z}}{\hbox{\cmss Z\kern-.4em Z}}
{\lower.9pt\hbox{\cmsss Z\kern-.4em Z}} {\lower1.2pt\hbox{\cmsss
Z\kern-.4em Z}} \else{\cmss Z\kern-.4em Z}\fi}
\def\II{\relax{\rm I\kern-.18em I}}
\def\IIa{{\II a}}
\def\IIb{{\II b}}
\def\IX{{\bf X}}
\def\ttb{Type $\II$B string theory}
\def\ndt{{\noindent}}
\def\bx{{\bf G}}
\def\sssec#1{\ndt$\underline{#1}$}
\def\CA{{\cal A}}
\def\CB{{\cal B}}
\def\CC{{\cal C}}
\def\CD{{\cal D}}
\def\CE{{\cal E}}
\def\CF{{\cal F}}
\def\CG{{\cal G}}
\def\CH{{\cal H}}
\def\CI{{\cal I}}
\def\CJ{{\cal J}}
\def\CK{{\cal K}}
\def\CL{{\cal L}}
\def\CM{{\cal M}}
\def\CN{{\cal N}}
\def\CO{{\cal O}}
\def\CP{{\cal P}}
\def\CQ{{\cal Q}}
\def\CR{{\cal R}}
\def\CS{{\cal S}}
\def\CT{{\cal T}}
\def\CU{{\cal U}}
\def\CV{{\cal V}}
\def\CW{{\cal W}}
\def\CX{{\cal X}}
\def\CY{{\cal Y}}
\def\CZ{{\cal Z}}
\def\B#1{{\bf #1}}
\def\BH{\B{H}}
\def\BM{\B{M}}
\def\BN{\B{N}}
\def\Ba{\B{a}}
\def\Bb{\B{b}}
\def\Bk{\B{k}}
\def\Br{\B{r}}
\def\p{\partial}
\def\pb{\bar{\partial}}
\def\eV{{\rm eV}}
\def\TeV{{\rm TeV}}
\def\GeV{{\rm GeV}}
\def\cm{{\rm cm}}
\def\dir{{\CD}\hskip -6pt \slash \hskip 5pt}
\def\dd{{\rm d}}
\def\Dslash{\rlap{\hskip0.2em/}D}
\def\cb{\bar{c}}
\def\ib{\bar{i}}
\def\jb{\bar{j}}
\def\kb{\bar{k}}
\def\lb{\bar{l}}
\def\mb{\bar{m}}
\def\nb{\bar{n}}
\def\ub{\bar{u}}
\def\wb{\bar{w}}
\def\sb{\bar{s}}
\def\tb{\bar{t}}
\def\vb{\bar{v}}
\def\xb{\bar{x}}
\def\zb{\bar{z}}
\def\Cb{\bar{C}}
\def\Db{\bar{D}}
\def\Tb{\bar{T}}
\def\Zb{\bar{Z}}
\def\half{{1\over 2}}
\def\bra#1{{\langle}#1|}
\def\ket#1{|#1\rangle}
\def\bbra#1{( #1|}
\def\kket#1{|#1 )}
\def\vev#1{\langle{#1}\rangle}
\def\codim{{\mathop{\rm codim}}}
\def\Hol{{\rm Hol}}
\def\cok{{\rm cok}}
\def\rank{{\rm rank}}
\def\coker{{\mathop {\rm coker}}}
\def\diff{{\rm diff}}
\def\Diff{{\rm Diff}}
\def\Tr{{\rm Tr~}}
\def\tr{{\rm tr~}}
\def\Id{{\rm Id}}
\def\vol{{\rm vol}}
\def\Vol{{\rm Vol}}
\def\c{\cdot}
\def\sdtimes{\mathbin{\hbox{\hskip2pt\vrule height 4.1pt depth -.3pt
width.25pt\hskip-2pt$\times$}}}
\def\dim{{\rm dim~}}
\def\Re{{\rm Re~}}
\def\Im{{\rm Im~}}
\def\imp{$\Rightarrow$}
\def\danger{{NB:}}
\def\Lie{{\rm Lie}}
\def\mod{{\rm mod}}
\def\lieg{{\underline{\bf g}}}
\def\liet{{\underline{\bf t}}}
\def\liek{{\underline{\bf k}}}
\def\lies{{\underline{\bf s}}}
\def\lieh{{\underline{\bf h}}}
\def\clieg{{\underline{\bf g}}_{\scriptscriptstyle{\IC}}}
\def\cliet{{\underline{\bf t}}_{\scriptstyle{\IC}}}
\def\cliek{{\underline{\bf k}}_{\scriptscriptstyle{\IC}}}
\def\clies{{\underline{\bf s}}_{\scriptstyle{\IC}}}
\def\CCK{K_{\scriptscriptstyle{\IC}}}
\def\inbar{\,\vrule height1.5ex width.4pt depth0pt}
\def\IIa{IIa}
\font\cmss=cmss10 \font\cmsss=cmss10 at 7pt
\def\sdtimes{\mathbin{\hbox{\hskip2pt\vrule height 4.1pt
depth -.3pt width .25pt\hskip-2pt$\times$}}}
\def\ao{{\bf a}}
\def\co{{\bf a}^{\dagger}}

\def\a{{\alpha}}
\def\ap{{l_s^2}}
\def\b{{\beta}}
\def\d{{\delta}}
\def\g{{\gamma}}
\def\e{{\epsilon}}
\def\z{{\zeta}}
\def\ve{{\varepsilon}}
\def\vf{{\varphi}}
\def\kk{{\kappa}}
\def\m{{\mu}}
\def\n{{\nu}}
\def\u{{\Upsilon}}
\def\l{{\lambda}}
\def\s{{\sigma}}
\def\t{{\theta}}
\def\o{{\omega}}
\def\seealso{{See also }}
\def\hepth#1{}
\def\hepph#1{}
%
\title{Physics of String Flux Compactifications}

\author{Frederik Denef
 \affiliation{Institute for Theoretical Physics, University of Leuven, Belgium}
 Michael R. Douglas
 \affiliation{Department of Physics and Astronomy, Rutgers University\\
 Institut des Hautes Etudes Scientifiques, Bures-sur-Yvette, France}
 Shamit Kachru
 \affiliation{Department of Physics and SLAC, Stanford University}
}

\begin{abstract}
We provide a qualitative review of flux compactifications of string
theory, focusing on broad physical implications and statistical
methods of analysis.

\end{abstract}

\maketitle



\section{Introduction}
\label{sec:intro}

String theory was first proposed as a candidate theory of quantum
gravity in 1974 \cite{Scherk:1974mc}.  Over the subsequent years,
supersymmetric versions of the theory were developed, and arguments
made that they were perturbatively finite
\cite{Schwarz:1982jn}.
The discoveries of anomaly cancellation
\cite{Green:1984sg}
and of quasi-realistic compactifications of the heterotic string
\cite{Candelas:1985en} in 1984--85, a period sometimes
referred to as the ``first superstring revolution,'' led very
rapidly to a broad consensus among particle physicists that
superstring theory was a viable contender for a ``theory of
everything,'' describing all of fundamental physics.

Now at this point, there were several good reasons to be wary of
this claim.  One was that there appeared to be many competing
theories; besides the different varieties of superstrings, there was
eleven-dimensional supergravity.  Another was the fact that string
theory was defined only as a perturbative expansion, which could
only be used directly at weak coupling.  Besides the evident fact
that the real world includes strongly coupled QCD, various arguments
had been made that constructing a completely realistic model would require
non-perturbative physics, most notably at the stage of supersymmetry
breaking
\cite{Dine:1985he}.

These doubts were addressed in a rather striking way in the ``second
superstring revolution'' of 1994--97, in which it was convincingly
argued that all of these theories are limits or aspects of one
unified framework, usually now called ``string/M theory.''  The
central idea, called duality, is that the strong coupling limit of
one of the various string theories, can be equivalent (or dual) to
another weakly coupled theory, one of the strings or else M theory,
the eleven-dimensional limit
\cite{Sen:2001di}.
Besides providing a variety of
nonperturbative definitions for string/M theory, these ideas also
led to exact solutions for the effective Lagrangians of a variety of
supersymmetric field theories, realizing phenomena such as
confinement and chiral symmetry breaking
\cite{Intriligator:1995au}.
Thus, while much remained
to do, it now seemed reasonable to hope that with further
theoretical progress, the remaining gaps such as supersymmetry
breaking could be addressed.

However, although the particle physics side of the story was moving
along nicely, serious gaps remained on other fronts, especially
cosmology.  The well known cosmological constant problem remained
\cite{Weinberg:1988cp,Weinberg:2000yb},
as well as the question of whether and how inflation could be
described. On a different level, ever since the first studies of
string compactification, it had appeared that making any concrete
construction required making many arbitrary choices, such as the
choice of extra dimensional manifold, a choice of gauge bundle or
brane configuration, and so on.  Other than the evident constraint
that the correct choice should lead to physics which fits the
observations, no principles had been proposed, even speculative
ones, that would suggest that any of these choices were preferred.
Thus it was quite unclear how to get testable predictions from the
theory.

A smaller scale version of this problem is that the metric in the
extra dimensions, as well as the other data of a solution of string
theory, depends on continuous parameters, called moduli.  Thus, even
having chosen a particular extra dimensional manifold, one has many
continuous parameters which enter into observable predictions.
Again, no particular values of these parameters appeared to be
preferred.

At first, one might compare this ambiguity with the choice of
coupling constants in a renormalizable field theory.  For example,
the Standard Model has $19$ parameters, and consistency conditions
only lead to very weak constraints on these, such as the unitarity
bound on the Higgs mass
\cite{Chanowitz:1988ae}.
However, the situation
here is essentially different -- string/M theory has no free
parameters, rather each of the parameters of a solution corresponds
to a scalar field in four dimensions.  If this choice is
unconstrained by the equations of motion, this implies that the
scalar field is massless.  And,
massless scalar fields typically (though not always)
lead to modifications
to the gravitational force law, which are not observed
\cite{Adelberger:2003zx}.
Thus, this is a phenomenological problem, usually called the problem
of ``moduli stabilization,'' which must be solved to
get realistic models.

Now, having stated the problem, there is a simple argument for why
it will generally solve itself in realistic models, as follows.  The
vacuum structure of a field theory is governed by the effective
potential $V_{eff}(\phi)$, which incorporates all classical and
quantum contributions to the potential energy, such as Casimir
terms, loop corrections, instantons and the like.  These corrections
can in principle be computed from the bare Lagrangian, and in almost
all cases where this has been done take generic order one values (in
units of some fundamental scale) which respect the symmetries of the
bare Lagrangian.  In particular, this includes the masses of scalar
fields. Thus, massless scalar fields never appear in practice,
unless they are Goldstone bosons for a continuous symmetry, or
unless we can tune parameters (as is done in condensed matter
systems to approach critical points).

A noteworthy exception is found in supersymmetric field
theories, in which non-renormalization theorems preclude corrections
to the superpotential to all orders in perturbation theory.  While
some theories admit non-perturbative corrections, others do not, and
in those theories moduli are natural.  However, a realistic model
must break supersymmetry at some scale $M_{\rm susy} \ge 1 \TeV$,
and below this scale the previous argument applies.  Thus, the
moduli will ultimately gain masses of order $m \sim c M_{susy}$,
where $c$ is often a small number of the order $M_{susy}/M_{P}$.

This solution turns out to be problematic \cite{Banks:1993en,
deCarlos:1993jw}, as it leads to the so-called Polonyi problem
\cite{Coughlan:1983ci}, wherein the light moduli fields carry too
much energy in the early universe, leading to overclosure.  Thus,
one is led to look for other mechanisms which could give larger
masses to moduli.  A particularly simple mechanism, which we discuss
in some detail in section 2, is to postulate a background
generalized magnetic field in the extra dimensions, usually called
``flux.''  The energy of such a field will depend on the moduli, and
provides a new contribution to the effective potential.  Its scale
is set by the unit of quantization (a fundamental scale such as the
string scale) and the inverse size of the extra dimensions.  This is
typically far higher than $M_{susy}$ and solves the problem.

But, there is an unexpected side effect of this mechanism.  On a
qualitative level, it works for fairly generic nonzero choices of
flux.  Supposing that each flux can take of order $10$ values, then
one finds of order $10^K$ distinct solutions, where $K$ is the number
of distinct topological types of flux (usually a Betti number).  And,
the compactification manifolds used in string theory, say the
Calabi-Yau manifolds, typically have $K \sim 30-200$.  Thus,
one finds a large vacuum multiplicity.  Furthermore, physical
predictions depend on this choice, both directly and because the
values of the stabilized moduli depend on the flux.

As one might imagine and as we discuss below, this vacuum
multiplicity makes it rather complicated to propose definitive tests
of the theory.  Philosophically, it seems very much at odds with the
idea that a fundamental theory should be simple and unique.  Now it
is hard to know how much weight to put on such considerations, which
do not in themselves bear on the truth or falsity of the theory as a
description of nature, but certainly one should ask for more
evidence before accepting this picture.

In a seemingly unrelated development, starting in the late
1990's, convincing evidence has accumulated for a non-zero dark
energy in our universe, from the accelerated expansion as measured
by observations of supernova, in precision measurements of the CMB,
and from other sources
\cite{Peebles:2002gy}.
Although not absolutely proven, the
simplest model for this dark energy is a positive cosmological
constant, of order $\Lambda = (0.71\pm 0.01)\Omega$, where $\Omega$
is the critical density.

This observation brought the cosmological constant problem to the
fore of fundamental physics.  As long as the data was consistent
with $\Lambda \sim 0$, the simplest hypothesis was that the correct
theory would contain some mechanism which adjusted $\Lambda$ to
zero.  Many proposals along these lines have been made, and
although none is generally accepted, nor is there any proof that
this is impossible. On the other hand, the existing proposals
generally do not fit well within string theory, and it had been
widely felt that a proper solution would require some essentially
new idea.

On the other hand, the problem of fitting a specific non-zero
$\Lambda$ looks rather different, indeed the particular choice of
value seems to be rather arbitrary.  Its one evident property is
that it is of the same order as the matter density, but only at the
present epoch (the coincidence problem).  This led to the study of
proposals such as quintessence in which the dark energy is time
dependent; we will only say here that this proposal has the
advantage of being directly testable, and so far the evidence seems
to be coming down on the other side, for a fixed cosmological
constant.

Now, there was one previous proposal for a mechanism which could
explain a small non-zero cosmological constant, due to
Banks \cite{Banks:1984tw}, Linde
\cite{Linde:1984ir} and particularly Weinberg
\cite{Weinberg:1987dv}.
This was the idea that the
underlying theory might manifest a large number of distinct vacua,
which are stable at least on cosmological time scales, and realize
different values of the cosmological term $\Lambda$.
When Weinberg was writing, the
direct observational bounds on $\Lambda$ placed a bound on its value
which had just excluded the range of $\Lambda$s which are too large to
allow successful galaxy formation.  Weinberg advocated the point of
view that given that $\Lambda$ is absurdly small compared to
fundamental scales in Nature (indeed, at the time he wrote, there was
only a tiny upper bound on its value), one should postulate that the
distribution of $\Lambda$ in the different vacua is ${\it uniform}$ in
the observationally allowed range, that compatible
with galaxy formation.  It then follows statistically that most
vacua which are compatible with observations would lie within a decade
or two of the maximal allowed value.  Thus, Weinberg predicted
that the eventual observed $\Lambda$ would be comparable
to the maximal value compatible with galaxy formation.  This has
proved true.

An important ingredient in this argument is that the fundamental
underying theory should admit many vacua.  While from the beginning,
the choices present in string compactification made it clear that the
theory has many distinct solutions, the number required by Weinberg's
argument (roughly $10^{120}$) was far larger than any reliable
approximation of the number of metastable vacua. Though some estimates
of very large numbers of constructions were made by counting distinct
soluble points (lattice constructions or orbifold models) in
continuous moduli spaces of vacua
\cite{Lerche:1986cx},
it was widely believed and still
seems very likely to be true that these many constructions would
collapse into a few after supersymmetry breaking, since they were
simply distinct points on the same supersymmetric moduli space.  The
number of discretely differing solutions seemed far smaller: for
instance, while Calabi-Yau spaces are far from unique, the number of
distinct choices after more than twenty years of systematically
searching for constructions still numbers in the thousands.  This
fact, together with the inability to exhibit reliable metastable
solutions after supersymmetry breaking, largely discouraged
speculation on this topic.

The duality revolution, however, brought many new ingredients of string theory
to the fore.  One of these was the generalized gauge fields which couple to
D-branes.  Each of these gauge fields come with an associated field strength,
whose flux can thread the extra dimensions.  It was realized in
\cite{Bousso:2000xa}
that in compact manifolds with sufficiently complicated topology, the
number of flux choices allowed by the known constraints could easily
exceed $10^{120}$.  This provided a large, discretely varying
collection of models, which could realize different values of the
cosmological term.
While this model neglected many aspects of the physics, in particular
simply freezing the moduli of the internal space by hand,
it made it very plausible
that string theory could contain the large set of
vacua required by Weinberg's argument.

In light of
this combination of theoretical and observational considerations,
it has become very interesting to find concrete and computable
models of string compactification in which the moduli problem is
solved, and in which distinct choices of flux provide a large collection
of vacua.  Starting around 2000, this problem received a great
deal of attention (some representative
works being \cite{Gukov:1999ya,Dasgupta:1999ss,Mayr:2000hh,Curio:2000sc,
Giddings:2001yu,Silverstein:2001xn}),
with
a milestone being the KKLT
proposal \cite{Kachru:2003aw} for a class of IIb compactifications in
which all moduli are stabilized, by a combination of fluxes and
non-perturbative effects, while supersymmetry is broken with a small
positive cosmological constant.  Subsequent work has proposed more
and more concrete realizations of this proposal \cite{Denef:2004dm,
Denef:2005mm,Lust:2006zg}, as well as
alternate constructions using the other string theories and M
theory, or relying on different hypotheses for the scales and
dominant terms in the effective potential.
One can (heuristically) envision these potentials as
taking values over a large configuration space,
which has been called the ``string landscape''
\cite{Susskind:2003kw}.
A central question in recent years has
been to characterize the landscape, and find dynamical mechanisms for
populating and selecting among the vacua within it.

A full discussion of these constructions very quickly becomes
technical, and we refer to reviews such as
\cite{Douglas:2006es,Blumenhagen:2006ci,Grana:2005jc,
Silverstein:2004id, Frey:2003tf}
for the details.  However, we will outline one
representative set of constructions in the next section, to illustrate the
ideas.

In the rest of the review, we begin to address the questions of
testability raised by the landscape.  In section 3, we discuss
formal results for the distribution of vacua and their observable
properties, while in section 4, we survey the various approaches for
getting concrete and testable predictions from the framework.

Many related topics had to be omitted for reasons of space, such as
inflation in string theory.  We were also limited to citing only major
reviews and a few influential and/or particularly recent papers, and
again direct readers to
\cite{Douglas:2006es,Blumenhagen:2006ci,
Grana:2005jc,Silverstein:2004id, Frey:2003tf}
for a more complete bibliography.

\section{Examples of flux vacua}
\label{sec:example}

In this section, we describe in more detail the ingredients that enter
in string flux compactifications.  We first give a simple
toy model that captures much of the relevant physics.  In \S2.2, we
extend this to a full discussion of a simple class of flux vacua of
IIa string theory.  In \S2.3, we briefly describe other ingredients
that enter in making quasi-realistic models.

\subsection{A toy model}

The essential point will be that fluxes and branes present in 10d
string theory provide natural ingredients for stabilizing the moduli
of a compactification.  However the basic ideas are independent of dimension
and the
simplest illustration involves 6d Einstein-Maxwell theory.  We
imagine, therefore, a 6d theory whose dynamical degrees of freedom
include the metric $g_{MN}$ and an abelian Maxwell field $F_{MN}$.
The Lagrangian takes the form
\begin{equation}
L = \int d^6x \sqrt{-g}\left( M_{6}^4 {\cal R} - M_{6}^2 |F|^2 \right)
\end{equation}
where $M_6$ is a fundamental unit of mass (the 6d Planck scale).

To define a compactification of this 6d theory to four dimensions, one
must choose a compact 2d internal space.  The topologies of
2d manifolds without boundary are classified by the genus, {\it i.e.},
the possible spaces $M_g$ are (hollow) donuts with $g=0,1,2,\cdots$ holes.
$g=0$ is the sphere, $g=1$ is the two-torus, and so forth.


Suppose we compactify the theory on the manifold $M_g$ of genus $g$
and total volume $R^2$.  As an ansatz for the 6d metric, we could take
\begin{equation}
\label{ansatz}
ds^2 = g_{\mu\nu} dx^{\mu}dx^{\nu} + R^2 \tilde g_{mn}dy^m dy^n
\end{equation}
with $\tilde g$ a metric of unit volume on $M_g$.  Our focus here will
be the dynamics of the ``modulus field" $R(x)$ which enters into the
ansatz Eq. (\ref{ansatz}).

We can expand the resulting 4d effective Lagrangian in powers of derivatives, as
is standard in effective field theory.  The leading terms arising from the
gravitational sector are
\begin{equation}
\label{gravitysec}
M_6^4 R^2 \int d^4x \sqrt{-g} \left( [ \int d^2y \sqrt{\tilde g}{\cal R}_2]  +
R^2 {\cal R}_4 \right) + \cdots
\end{equation}
Here, ${\cal R}_2$ is the curvature of $\tilde g$ and ${\cal R}_4$ is
the curvature of the 4d spacetime metric $g$.  The $\cdots$ includes
both gradients of $R$ and terms involving the 4d gauge field.

There are two important points about the action as written above.
Firstly, defining $M_4^2 = M_6^4 R^2$ as the 4d Planck scale, we see the action
is not in Einstein frame -- the kinetic terms of $R$ and the graviton are mixed.
Secondly, the first term in brackets above is in fact a topological invariant,
$\chi(M_g)$.  For a surface of genus $g$, $\chi = 2-2g$.  We therefore find a
scalar potential for the modulus field unless $g = 1$.

To go to 4d Einstein frame, we should redefine the 4d metric: $g \to h
= R^2 g$.  Then $\sqrt{h}{\cal R}_{h} = R^2 \sqrt{g}{\cal R}_g$, and
we find a 4d Lagrangian
\begin{equation}
\label{gravweyl}
M_4^2 \int d^{4}x \sqrt{-h}\left( {\cal R}_h - V(R) \right)
\end{equation}
The potential $V(R)$ is given by
\begin{equation}
\label{curvpot}
V(R) \sim (2g-2) {1\over R^4}~.
\end{equation}
The $R$ dependence just follows from the Weyl rescaling above.

We have learned an interesting lesson.  In the absence of background
Maxwell fields, of the infinite family of compactification topologies
parametrized by $g$, precisely one choice yields a static solution.
The choice $g=1$, the two-torus, yields a vanishing potential and a 4d
flat-space solution.  The positive curvature $g=0$ surface, the
two-sphere $S^2$, naively has a negative potential $\sim -{1\over
R^4}$, yielding runaway to $R\to 0$ (where the regime of
trustworthiness of our analysis breaks down).  The negative curvature
options, $g>1$, instead yield positive potentials, which vanish as $R
\to \infty$ -- they spontaneously ``decompactify'' back to the 6d flat
space solution.

All of these features of our toy model
have analogues in full-fledged string compactification.
The $T^2$ here, which is Ricci-flat, is analogous to the Calabi-Yau
solutions which have been intensely studied by string theorists.
One difference is that while $T^2$ is a unique 2-manifold, there are
at least thousands of topologically distinct Calabi-Yau threefolds, each
leading to a class of compactifications.
If our 6d theory were promoted to a supersymmetric Einstein-Maxwell theory,
this 4d flat solution would have unbroken supersymmetry at the
KK scale (which partially explains the
fascination that string theorists have had with the analog
Calabi-Yau solutions).
Notice that the modulus field $R$ survives ``unfixed'' in the 4d effective
theory -- it appears to 4d physicists as a massless scalar field.
In fact the $T^2$ has an additional modulus corresponding to its shape;
the Calabi-Yau spaces also manifest both volume
moduli and ``shape'' (complex structure) moduli.

The $S^2$ case, with positive curvature, mirrors the Einstein
manifolds used in Freund-Rubin compactification \cite{Freund:1980xh}
-- more on this momentarily.  The negative curvature manifolds (here,
the most common topologies) are just now being examined as candidate
string compactifications \cite{McGreevy:2006hk}.

\subsubsection{Including fluxes and branes}

Let us do this step by step, beginning by turning on the Maxwell
field.  To find 4d vacuum solutions with maximal symmetry, we should
only allow the gauge field configuration to be  nontrivial on $M_g$.  The most
obvious possibility is to thread $M_g$ with some number of units of
magnetic flux of the gauge field
\begin{equation}
\int_{M_g} F = n~.
\end{equation}
Like a magnetic flux threading a solenoid, this flux through $M_g$
carries positive energy density.  This makes an additional
contribution to the 4d effective potential as a function of $R$.
Naively, it scales like ${1\over R^2}$ -- a factor of $R^2$ from the
determinant of the metric, and two factors of $1/R^2$ from the metric
factors contracting the indices on $F_{mn}$.  However, upon rescaling
to reach 4d Einstein frame, one multiplies by an additional $1/R^4$,
so the full potential takes the schematic form
\begin{equation}
\label{curvflux}
V(R) \sim (2g-2){1\over R^4} + {n^2\over R^6}~.
\end{equation}
In the presence of $n$ units of magnetic flux, the status of the different
compactification topologies changes.

Previously, the $g=0$ $S^2$ compactification yielded solutions that run to
(uncontrolled) small values of $R$.  However, it is easy to see that with
$n$ units of flux, the positive flux energy can balance the negative curvature
contribution at a finite values of $R$, which grows with $n$.  Therefore,
for large $n$, one obtains reliable flux vacua from a large $S^2$ with $n$
units of flux piercing it.  These are the Freund-Rubin solutions.

For $g=1$, where previously there was a moduli space of solutions,
the flux causes a runaway to large $R$.
Finally, for $g>1$, since both terms in the potential are positive and vanish
at large $R$, there is also a runaway.

The same behavior obtains in the full string theory.  The
Freund-Rubin solutions also exist there, and play a crucial role in
e.g. the AdS/CFT correspondence.  There are indeed no solutions
of string theory which start from (supersymmetric) Calabi-Yau
compactification and incorporate only flux -- additional
ingredients are needed (which we will come to next).  Finally, the
negative curvature spaces also require additional ingredients to avoid
runaway.

Next, we add ${\it branes}$, dynamical, fluctuating lower-dimensional
objects which can carry gauge and matter fields.  A common example is
the Dirichlet (or D) brane.  These have positive tension and thus make
positive contributions to $V(R)$.  String theories also include fixed,
non-dynamical objects, the so-called orientifold planes or O-planes,
with negative tension.

We now modify our toy model to include some fixed ${\cal O}(1)$ number $m$ of
O3-planes in the background, at points on $M_g$.
Taking into account Weyl rescaling, their contribution
to the 4d effective potential is
\begin{equation}
\label{vor}
\delta V_{O3} = - m {1\over R^4}
\end{equation}

We see that after inclusion of these objects, the full effective potential of
our toy model takes the general form
\begin{equation}
\label{vfull}
V(R) = (2g-2){1\over R^4} - m{1\over R^4} + n^2 {1\over R^6}~.
\end{equation}
The physics of Eq. (\ref{vfull}) captures many of the qualitatively
important facts in flux compactification (though the identical scaling
of the O-plane and curvature contributions will not hold in the string
analogues).  Of course it is an oversimplification: in real
constructions there are many further constraints, determining the
number of branes, planes and so forth.

Including both branes and fluxes, we see that if models with suitable $m$
are available, (a) the modulus $R$ of the $T^2$ compactification can be fixed
by inclusion of O3 planes, and (b) the negative curvature models may now admit
critical points of their potential at large $R$ (for string theory
examples, see \cite{Saltman:2004jh}).

The vacua analogous to type (a) will be our focus in much of this
review.  These are Calabi-Yau models with an underlying ${\cal N}=2$
supersymmetry, broken to ${\cal N}=1$ by the branes and O-planes.
Other ingredients, such as non-perturbative physics, are required
to stabilize all moduli and break supersymmetry,
as described in  \cite{Douglas:2006es}.

In many of these models, the scale of supersymmetry breaking can be
very low compared to the KK scale.  Thus these models are a logical
place to search for string extensions of the MSSM, which include its
coupling to quantum gravity.

\subsection{IIa flux vacua}

The simplest full constructions of string vacua along these lines
use IIa string theory
\cite{DeWolfe:2005uu,Villadoro:2005cu,Camara:2005dc}.
This theory contains p-form
RR field strengths $F_p$ with $p=0,2,4,6,8$, and
an NS field strength $H_3$.  Practically, this
means that in addition to the possible fluxes threading nontrivial
0,2,3,4 and 6 cycles of the 6-dimensional compactification manifold,
the theory contains
D0,2,4,6,8 branes and NS 5 branes.  Along with the D-branes,
there can be O-planes of each even dimension $<10$.  Any of these objects
can form part of a solution with 4d rotational
symmetry if they ``fill'' 4d space-time,
and thus ``wrap'' a $p-3$ dimensional cycle in the compactification
manifold.

We can analyze the resulting effective potential in close analogy to
the one for our toy model.  Let us imagine compactification on a
Ricci-flat 6-manifold $M$ (say a Calabi-Yau) with $b_{2}=b_{4}$
topologically distinct 4-cycles, $b_3$ distinct 3-cycles, no 1-cycles
or 5-cycles, and volume $R^6$.

Unlike our toy model, in string theory there is no predetermined
dimensionless coupling; the 10d coupling $g_s$ is determined by the
dilaton field $\phi$ as $g_s = e^{\phi}$.  A minimal discussion
therefore focuses on the dynamics of the volume modulus $R$ and
$\phi$, and we shall do that below, sketching the elaboration to
include other moduli at the end.  In a full discussion, all moduli
must be fixed.

Because the manifold is Ricci-flat, there is no contribution to the
potential from the 10d curvature term.  Thus, to prevent a runaway to
large $R$, we will need to include O-planes.  Given the assumed
topology of $M$, these must be O6-planes wrapping 3-cycles.

We will also need to know the $g_s$ scaling of the various sources of
energy in ten dimensions.  In 10d string frame, the Einstein term
scales like $e^{-2\phi}$, while the D-brane/O-plane tensions scale
like $e^{-\phi}$.  In addition, the RR $|F|^2$ terms are independent
of $g_s$ (in the convention where the RR flux quantization is $g$
independent), while the NS $|H|^2$ arises at order $e^{-2 \phi}$.

We can now write a schematic effective potential
for $R$ and $e^{\phi}$, given any particular choice of ingredients.  One
natural class of compactifications involves $N$ unit of RR $F_4$ flux,
${\cal O}(1)$ units of $F_0$ flux and $H_3$ flux, and some fixed ${\cal O}(1)$
number of O6 planes.   The resulting potential is
\begin{equation}
\label{Vtwoa}
V = N^2 {e^{4\phi}\over R^{14}} + {e^{2\phi}\over R^{12}} - {e^{3\phi}\over R^9}
+ {e^{4\phi}\over R^6}~.
\end{equation}
One sees that this potential admits vacua with $R \sim N^{1/4}$ and
and $g_s = e^{\phi} \sim N^{-3/4}$.  So by analogy with the
Freund-Rubin toy model, this theory has vacua with large volume (and
weak coupling) if one turns on a large number of flux quanta.  An
important distinction between this model and the Freund-Rubin vacua is
that the size of the compact space $M$ is parametrically smaller, in
this leading approximation, than the curvature radius set by the
leading estimate for the 4d cosmological term.  So these models are
indeed models of 4d matter coupled to quantum gravity, over some range
of scales.

This discussion has clearly oversimplified the complex problem of
stabilizing Calabi-Yau moduli.  However, the basic construction above
does in fact yield models with all moduli stabilized in the full
string theory.  The simplest example \cite{DeWolfe:2005uu}
uses the Calabi-Yau orbifold $T^6/Z_3^2$.
This space is a quotient of the six-torus by the action of two discrete
symmetry groups.  We start with the torus described by complex
coordinates $z_i = x_i + i y_i$, $i=1,..,3$, subject to
identifications
\begin{equation}
z_i \sim z_i + 1 \sim z_1 + \alpha
\end{equation}
where $\alpha = e^{i\pi/3}$.  This torus has a $Z_3$ symmetry $T$ under which
\begin{equation}
T:~(z_1,z_2,z_3) \to (\alpha^2 z_1, \alpha^2 z_2, \alpha^2 z_3)~.
\end{equation}
This action has a total of $3\times 3 \times 3 = 27$ fixed points on
the torus.
While these result in conical singularities, these are allowed in
perturbative string theory \cite{Dixon:1985jw}.  One can
reduce this to $9$ singularities by
quotienting by another, freely acting $Z_3$ symmetry
$Q:~(z_1,z_2,z_3) \to (\alpha^2 z_1
 + {{1+\alpha}\over 3}, \alpha^4 z_2 + {{1+\alpha}\over
3},z_3 + {{1+\alpha}\over 3})$.

Orbifolds by discrete subgroups of $SU(3)$ give rise to (in general
singular) Calabi-Yau manifolds, so our construction thus far gives an
${\cal N}=2$
type IIa compactification on a Calabi-Yau space.
To break the
supersymmetry to ${\cal N}=1$, we now ${\it orientifold}$ by the
simultaneous action of the $Z_2$ involution
\begin{equation}
\sigma: ~z_i \to - \overline{z_i}
\end{equation}
with worldsheet parity reversal.  The action of $\sigma$ gives rise to
a fixed 3-plane $z_i=0$, which is wrapped by a space-filling O6 plane.

The model we have described has 12 volume moduli, whose origin is as
follows: three of them simply rescale the volumes of the three
two-tori involved in the construction.  The other 9, one per
singularity, are associated to ``exceptional cycles''
introduced by resolving the
$C^3/Z_3$ fixed points.  On the other hand, it has no
complex structure moduli (it is ``rigid''), because the $Z_3$ symmetries
only arise for a unique shape of the covering torus, and because
resolving the singularities introduces only volume moduli.

It is now straightforward to add ingredients which give rise to a
potential analogous to that in Eq. (\ref{Vtwoa}).  We already
have an O6 plane; one can thread the complementary
three-cycle with $H_3$ flux, and turn on $F_0$ flux and $F_4$ flux through the
4-cycles transverse to the $T^2$s of the original $(T^2)^3$.  The one
novelty is the existence of the exceptional cycles, but these can
be stabilized at large volume by four-form flux as well.
The details can be found in \cite{DeWolfe:2005uu};
the result is a model in which the leading effective potential
stabilizes the moduli at weak string coupling and large volume,
just as suggested by Eq. (\ref{Vtwoa}).  It should be admitted
that this analysis is not an
absolute proof of existence; for possible subtleties see \cite{Banks:2006hg}.

\subsection{More realistic models}

We have focused here on construction of models with computable moduli
potentials.  To make contact with observed physics, we also need
a sector giving rise to (a supersymmetric extension of)
the Standard Model, and perhaps a sector which is responsible for
supersymmetry breaking and its transmission to the Standard Model.
There has been significant progress on constructing flux vacua which
incorporate all of these elements.

One popular method for constructing Standard-like models in string
theory is to use intersecting D-branes, for example D6 branes in IIa
theory.  Stacks of $N$ parallel branes manifest an $SU(N)$ gauge
theory, while intersections of D6-branes at points in the extra
dimensions localize chiral matter multiplets \cite{Berkooz:1996km}.
These ingredients allow ``engineering'' the Standard Model, and
indeed fairly general ${\cal N}=1$ gauge theories.
For the state of the art, see \cite{Blumenhagen:2005mu}.

One can also use this to construct models of dynamical supersymmetry
breaking.  It has recently become clear that even the simplest
non-chiral gauge theories have supersymmetry breaking vacua
\cite{Intriligator:2006dd}, and one can very simply engineer these
constructions on D6 branes \cite{Ooguri:2006bg}.  Another option is to
arrange for the flux potential itself to provide the supersymmetry
breaking, as we will discuss in the next section.  Thus it is quite
plausible that the class of models we described above includes the
supersymmetric Standard Model and its various extensions, with
supersymmetry breaking transmitted via gravity or gauge
interactions.

\section{Statistics of vacua}
\label{sec:stat}

We now supply a brief introduction to and survey
of this topic, referring to
\cite{Douglas:2006es,Kumar:2006tn}
for more extensive reviews.

\subsection{The Bousso-Polchinski model}

As pointed out in \cite{Bousso:2000xa}, the freedom one has to turn on
various independent flux quanta in string theory compactifications
leads to ensembles of vacua with a variety or ``discretuum'' of low
energy effective parameters.  This leads to a need for statistical
analysis of the resulting vacuum distributions \cite{Douglas:2003um}.

This is true in particular for the cosmological constant, implying
naturally the existence of string vacua with exceedingly small
effective four dimensional cosmological constants, such as our own,
without the need to invoke any (so far elusive) dynamical mechanism
to almost-cancel the vacuum energy.

To see how this comes about, consider
the (classical) potential induced by a flux $F$ characterized by
flux quanta $N^i \in \IZ$, $i=1,...,K$, given by
\begin{equation}
 V_N(\phi) =
 V_0(\phi) + \int_X \| F \|^2 = V_0(\phi) + \sum_{i,j} g_{ij}(\phi) \, N^i
 N^j,
\end{equation}
where $\phi$ denotes the moduli of the compactification manifold $X$
and $g_{ij}(\phi)$ is some positive definite effective metric on the
moduli space. The number of fluxes $K$ is typically given by the
number of homologically inequivalent closed cycles of some fixed
dimension in $X$, which for the known examples of six dimensional
Calabi-Yau manifolds is typically of order a few hundred. The bare
potential $V_0(\phi)$ is taken to be negative. In string
compactifications it could e.g.\ come from orientifold plane
contributions, and in this context $V_0$ (as well as $g_{ij}$) will
be of order of some fundamental scale such as the string or
Kaluza-Klein scale.

Each vacuum of this model is characterized by a choice of flux
vector $N$ together with a minimum $\phi_*$ of $V_N(\phi)$. Finding
these critical points explicitly is typically impossible, so to make
progress one has to use indirect statistical methods, as we will
discuss further on. However, before plunging in to this full, coupled
problem, let us, following \cite{Bousso:2000xa}, first simply freeze
the moduli at some fixed value $\phi=\phi_0$ and ignore their
dynamics altogether. In that case it is easy to compute the
distribution of cosmological constant values: the number of vacua
with cosmological constant $\Lambda = V_N(\phi_0)$ less than
$\Lambda_*$ is then simply given by the number of flux lattice
points in a sphere of radius squared $R^2 = |V_0| + \Lambda_*$,
measured in the $g_{ij}$ metric. When $R$ is sufficiently large,
this is well-estimated by the volume of this ball, i.e. ${\rm
Vol}_K(R) = \frac{\pi^{K/2}}{\Gamma(1+K/2)} \frac{R^K}{\sqrt{\det
g}}$, leading to a $\Lambda$-distribution
\begin{equation} \label{BPdistr}
 d N_{\rm vac}(\Lambda) = \frac{\pi^{K/2}}{\Gamma(K/2)} \frac{\left(
 |V_0| + \Lambda
 \right)^{\frac{K}{2}-1}}{\sqrt{\det g}} \, d\Lambda \approx \left(
 \frac{2 \pi e \, (|V_0|+\Lambda)}{\mu^4} \right)^{K/2} \, \frac{d\Lambda}{|V_0|+\Lambda}.
\end{equation}
where $\mu^{4}:=(\det g)^{1/K}$ can be interpreted as the mass scale
of the flux part of the potential and we assumed large $K$ and used
Stirling's formula to get the last approximate expression. Note that
in particular at $\Lambda=0$, for say $|V_0|/\mu^4 \sim \CO(10)$, we
get a vacuum density $d N_{\rm vac} \sim 10^{K/2} \,
d\Lambda/|V_0|$. Hence for $K$ a few hundred, there will be
exponentially many vacua with $\Lambda$ in the observed range
$\Lambda \sim 10^{-120} M_p^4$, even if all fundamental scales
setting the parameters of the potential are of order $M_p^4$.

In such a model, there is no need to postulate either anomalously
large or small numbers, or an unknown dynamical mechanism, to obtain
vacua with a small cosmological constant.
Combined with a cosmological mechanism generating all possible vacua
(such as eternal inflation) and Weinberg's argument as discussed
in the introduction, we have a candidate solution to the
cosmological constant problem within string theory.

Of course, the Bousso-Polchinski model is only a crude approximation
to actual flux vacua of string theory; in particular freezing the
moduli at arbitrary values is a major
oversimplification. However, more general and more refined analyzes
of low energy effective parameter distributions of actual string
flux vacua taking moduli dynamics into account, initiated in
\cite{Douglas:2003um} and further developed in
\cite{Ashok:2003gk,Denef:2004ze,Denef:2004cf}, confirmed the general
qualitative features of the model we just discussed. We now turn to
an overview of these studies.

\subsection{A simple toy model}

Let us demonstrate the basic ideas behind these techniques by
considering the following ensemble of effective potentials:
\begin{equation} \label{toy1}
 V_{N,M}(\phi) = N \phi + M \frac{\phi^2}{2}, \quad N,M \in \IZ,
 \quad N^2 + M^2 \leq L.
\end{equation}
Here the $(N,M)$ (crudely) model fluxes, while $\phi$ models a
modulus field. To get a large number of vacua, we take $L$ to be
very large. Finding the critical points is trivial for this
ensemble: they are given by $\phi_* = -N/M$, and are stable iff
$M>0$. To find the distribution of vacua over $\phi$-space in the
large $L$ limit, we could in principle start from these explicit
solutions. However, a more elegant and powerful method, which is
also applicable to cases for which explicit solutions cannot be
found, and to ensembles of actual string theory flux vacua, goes as
follows.

The number of vacua $\phi_*$ in an interval $I$ is given by
\begin{equation}
 \CN_{\rm vac}(I) = \sum_{N,M} \int_I d\phi \, \delta(V'(\phi)) \, |V''(\phi)| \,
 \theta(V''(\phi)),
\end{equation}
where $\theta(x):=1$ if $x>0$, $\theta(x):=0$ if $x<0$. The
integrand $\delta(V') |V''|$ gives a contribution $+1$ for each
critical point in $I$, while $\theta(V'')$ restricts to actual
minima.

Now in the large $L$ limit, we can approximate the sum over $(N,M)$
by an integral, and write
\begin{equation} \label{vaccount}
 \CN_{\rm vac}(I) \approx \int_I d\phi \, \rho(\phi), \qquad
 \rho(\phi):=\int dN \, dM \, \delta(V'(\phi)) \,
 V''(\phi) \, \theta(V''(\phi)),
\end{equation}
where $\rho(\phi)$ can be interpreted as a vacuum number density on
moduli space. To evaluate the integral over $(N,M)$ at a given fixed
$\phi$, it is convenient to make the following linear change of
variables $(N,M) \to (v',v'')$:
\begin{equation}
 v' := V'(\phi) = N + M \phi, \quad v'':= V''(\phi) = M.
\end{equation}
This change of variables has Jacobian $=1$, and the integration
domain in the new variables is $L \geq N^2+M^2 = (v'-v'' \phi)^2 +
(v'')^2$. The integral is now trivially evaluated, yielding the
distribution
\begin{equation} \label{dens}
 \rho(\phi) = \frac{L}{2} \frac{1}{1+\phi^2}.
\end{equation}
Note that this integrates to at total number of vacua $\CN_{\rm
vac}(\IR) \approx \pi L / 2$ --- this is as it should, since the
total number of pairs $(N,M)$ in the ensemble is approximated by the
volume $\pi L$ of the region $N^2+M^2 \leq L$, each $(N,M)$ leads to
a unique critical point, and half of those are minima. The density
Eq. (\ref{dens}) confirms the intuitive expectation that ``most'' vacua
in this ensemble will be at order 1 values of $\phi$, but makes this
far more precise.

Let us next consider a more general ensemble of the form
\begin{equation}
 V_{N,M}(\phi) = N f(\phi) + M g(\phi),
\end{equation}
where $f$ and $g$ are arbitrary functions. Now it is no longer
possible to proceed by finding explicit solutions --- even finding a
single explicit solution will typically be out of reach even for
simple choices of $f$ and $g$. On the other hand, the previous
computation is straightforwardly extended to this case, for general
$f$ and $g$, resulting in a vacuum number density
\begin{equation}
 \rho(\phi) d\phi = \frac{L}{2} \frac{|f' g'' - f'' g'|}{(f')^2 + (g')^2}
 \, d\phi
 = \frac{L}{2} \, {\rm sign}\left(\frac{g'}{f'}\right)' \, d \arctan \left( \frac{g'}{f'} \right).
\end{equation}
This illustrates the power of statistical methods over explicit
constructions.

An interesting special case, which has an important counterpart in
actual string flux vacua, is obtained by setting $f(\phi)=\phi$,
$g(x)=\phi \log \phi - \phi$, $\phi>0$.
Potentials with similar structure appear naturally in string theory,
as we will discuss in more detail below.  The corresponding vacuum
density is
\begin{equation} \label{scalinv}
 \rho(\phi) d\phi = \frac{L}{2} \frac{1}{(1+ \log^2 \phi)}
   \, \frac{d\phi}{\phi} =
 \frac{L}{2} d \arctan \log \phi.
\end{equation}
Note that this distribution is approximately scale invariant, and
thus naturally allows hierarchically small (and large) vacuum values
of $\phi$ and therefore $V(\phi)$. For this particular ensemble, we
can also see this directly as the critical points can again be found
explicitly, namely $\phi_* = e^{-N/M}$.


\subsection{Ensembles of flux vacua in string theory} \label{sec:stringfluxvac}

The most studied and best understood ensemble of flux vacua is the
IIb ensemble, which arises by allowing two different kinds of fluxes
(RR and NSNS) to be turned on through nontrivial 3-cycles of a
Calabi-Yau compactification space $X$. There are two kinds of
geometric moduli, arising from complex structure (shape) and
K\"ahler (size) deformations of $X$. Besides these, there is a
universal modulus $\tau=C_0 + i/g_s$, where $C_0$ is the universal
axionic scalar of the IIb theory and $g_s$ is the string coupling
constant. The complex structure moduli and $\tau$ appear
nontrivially in the potential induced by the fluxes and as a result
are generically stabilized at tree level. The K\"ahler moduli on the
other hand do not but can under certain conditions be stabilized by
quantum effects \cite{Kachru:2003aw}. We will assume this is the
case and just freeze them in the following.

The potential for this ensemble has the standard $\CN=1$
supergravity form
\begin{equation}
 V_{N}(z,\bar{z}) = e^K \left( g^{a \bar{b}} D_a W_{N}
 \overline{D_b W_{N}} - 3
 |W_{N}|^2 \right)
\end{equation}
where $g_{a \bar{b}} = \partial_a \bar{\partial}_{\bar b} K$, $D_a =
\partial_a + \partial_a K$ and
\begin{equation}
 W_N(z) = \sum_{i=1}^K N^i \Pi_i(z), \qquad K(z,\bar{z}) = -\ln \left(
 Q^{ij} \Pi_i(z) \overline{\Pi_j(z)} \right).
\end{equation}
Here $z^a$ denotes the complex structure coordinates together with
$\tau$, $Q^{ij}$ is a known constant $K \times K$ matrix and
$\Pi_i(z)$ are certain complicated but in principle computable
holomorphic functions (the periods), which depend on the Calabi-Yau
$X$ at hand. Note that the potential is quadratic in the $N_i$, as
in the Bousso-Polchinski model. The number of flux quanta is $K=2 \,
b_3(X)$, where $b_3(X)$ is the third Betti number of $X$, i.e.\ the
number of homologically independent 3-cycles. For the quintic
Calabi-Yau, which can be described as the zero locus of any
homogeneous degree 5 polynomial in $\IC^5$ with points related by
overall complex rescaling identified, one has $b_3 = 204$.

Finally, there is a tadpole cancellation constraint on the fluxes, of
the form
\begin{equation} \label{constraint}
 \frac{1}{2} Q^{ij} N_i N_j \leq L
\end{equation}
where $L$ depends again on the compactification data but typically
ranges from $\CO(10)$ to $\CO(1000)$.

Supersymmetric vacua of this model are given by solutions of $D_a
W_N(z)=0$. Note that counting these amounts to a direct
generalization of the counting problem of our toy model, with the
periods $\Pi_i(z)$ generalizing $f(\phi)$ and $g(\phi)$. And indeed
it turns out to be possible to derive approximate distributions at
large $L$ using essentially the same ideas
\cite{Ashok:2003gk,Denef:2004ze}. The resulting
distribution\footnote{To be more precise, this distribution is
obtained by dropping the absolute value signs around the Jacobian
determinant generalizing $|V''(\phi)|$ in Eq. (\ref{vaccount}). As a
result some vacua will be counted negatively, hence the density is
an index density rather than an absolute density. This nevertheless
gives a good estimate of the actual density, and in any case a lower
bound. Absolute densities can be obtained as well, but are more
complicated.} of vacua over moduli space turns out to be
surprisingly simple:
\begin{equation} \label{adformula}
 d N_{\rm vac}(z) = \frac{(2 \pi L)^{K/2}}{(\frac{K}{2})! \,
 \pi^{K/4}}\,
 \det(R(z)+\omega(z) \, {\bf 1}),
\end{equation}
where $R(z) = \frac{i}{2} R^a_{bc\bar{d}} dz^c \wedge
d\bar{z}^{\bar{d}}$ is the curvature form of the metric $g_{a
\bar{b}}$ and $\omega(z) = \frac{i}{2} g_{a \bar{b}} dz^a \wedge d
\bar{z}^{\bar{b}}$ the K\"ahler form.

Note the close similarity of the $z$-independent prefactor with the
prefactor of the Bousso-Polchinski distribution Eq. (\ref{BPdistr}).
Essentially this arises because the constraint Eq. (\ref{constraint})
combined with the condition of supersymmetry roughly restricts the
fluxes to be contained in a sphere of radius proportional to
$\sqrt{L}$. As a result, we will get similarly huge numbers of
actual IIb flux vacua in string theory
--- for the models analyzed in detail in \cite{Denef:2004dm}, we find
$N_{\rm vac}({\rm total}) \sim 10^{307}, 10^{393},
10^{506}$.

The last of these figures is the justification for the often quoted
number $10^{500}$ as the total number of string vacua.  However, as
discussed in detail in \cite{Douglas:2006es}, there are many further
uncertainties even in counting the known classes of vacua, so this
famous number should not be taken too seriously.  Furthermore, the
number we are ultimately most interested in, namely that of vacua
similar to our own, is even less well understood at present; it could
still be the case that only a few vacua (or even none) fit all the
constraints implicit in the existing data.  The point rather is that
the problem of computing numbers of vacua with specific properties is
both mathematically well posed (at least, as much so as string/M
theory itself), and far easier than constructing the actual vacua.
Thus, we can expect steady progress in this direction, leading to
results of fairly direct phenomenological interest, as we will shortly
explain.

The distribution Eq. (\ref{adformula}) has some interesting structure.
It diverges (remaining integrable) when the curvature diverges,
which happens near so-called conifold degenerations of the
Calabi-Yau manifold $X$, corresponding to a 3-sphere in $X$
collapsing to zero size. Near this 3-cycle, the Calabi-Yau can be
described by an equation of the form $x_1^2 + x_2^2 + x_3^2 + x_4^2
= z$ in $\IC^4$, with the 3-cycle being the real slice through it
and $z=0$ the conifold point. The vacuum density near this point is
given by \cite{Denef:2004ze}:
\begin{equation} \label{conifdistr}
 dN_{\rm vac}(z)
 \sim \frac{d |z|}{|z| \log^2|z|^{-2}} \sim d \, (\log |z|^{-1})^{-1}.
\end{equation}
Note this is approximately scale invariant, naturally allowing
hierarchically small scales for $z$, similar to the toy model
distribution Eq. (\ref{scalinv}). This is no coincidence: near the
conifold point, there is a pair of periods $\Pi(z) \sim (z,z \log z
- z)$, just like in our toy model.

In the case at hand, small values of $z$ give rise to large warped
$AdS_5$-like throats in the compactification \cite{Giddings:2001yu},
which have a dual gauge theory description through the AdS-CFT
correspondence \cite{Klebanov:2000hb}.
In terms of the dual gauge coupling $g$, the distribution is simply
uniform $dN_{\rm vac} \sim dg^2$.
The enhancement of vacua at small $z$ may be of more than
academic interest:
such warped throats have many possible phenomenological uses.
They can provide a natural
embedding of the Randall-Sundrum scenario
\cite{Randall:1999ee} in string theory;
they give rise to natural models of warped (and hence exponentially
low-scale) supersymmetry breaking
\cite{Kachru:2002gs}; and some of the simplest inflationary scenarios in
string theory make use of such throats \cite{Kachru:2003sx}.

From this one-parameter distribution one can reasonably guess that
the majority of IIb flux vacua will contain such warped throats, by
combining the fact that some sizable fraction contains such a throat
in the 1-parameter case with the simple argument that the
probability of having \emph{no} such throats in an $n$-modulus case
can be expected to be roughly the $n$-th power of the probability of
having no such throats in the one modulus case, becoming small at
large $n$. Some more concrete evidence for this has been given in
\cite{Hebecker:2006bn}.

Distributions of other quantities besides the moduli have also been
obtained for IIb ensembles. For example the distribution of
cosmological constants of supersymmetric vacua near zero turns out
to be uniform: $d N_{\rm vac}(\Lambda) \sim \theta(-\Lambda) \,
d\Lambda \sim d|W|^2$, and the same is true for the string coupling
constant: $dN_{\rm vac}(g_s) \sim dg_s$. If the K\"ahler moduli are
stabilized according to the KKLT scenario, the distribution of KK
scales $M_{KK}$ is roughly given by $d N_{\rm vac}(M_{KK}) \sim
e^{-c M_{KK}^{-4}} d(M_{KK}^{-4})$ with $c$ some constant depending
on the compactification manifold. This shows that low KK scales,
(i.e.\ large extra dimensions) are statistically excluded in this
scenario.

Distributions for other ensembles such as M-theory or type IIa flux
vacua vacua have been worked out as well, and studies of
distributions of discrete D-brane data such as gauge groups and
matter content of intersecting brane models have been initiated. We
refer to \cite{Douglas:2006es} and references therein for more
details.

\subsection{Supersymmetry breaking scale}

\label{sec:susybreaking}

Perhaps most interesting from a phenomenological point of view is the
scale of supersymmetry breaking.  We would like to know, out of all the
string/M theory vacua which agree with existing data, is it likely that
we live in one with low scale breaking, leading to discovery of
superpartners at LHC, or not?

Of course, to answer this question conclusively, one would need a much
better handle on many issues, both affecting the distribution of vacua,
and also how cosmological selection mechanisms and the like influence
actual probabilities on the string theory landscape.  Since at present we
know very little about the second question, we will stick to the first,
and discuss the better defined number distributions of string flux vacua.

Let us say that if many more vacua have property ``$X$'' than property
``not $X$'', then $X$ is ``favored,'' at least within the
considerations we discuss.  Another term for this type of
consideration is ``stringy naturalness.''  Of course, if it were to
turn out that the probabilities of different vacua were roughly equal,
this would lead directly to a statistical prediction.  Even if not, if
the probabilities to obtain vacua were \emph{uncorrelated} with the
property of interest, one would also get a statistical prediction.  On
the other hand, a particular cosmological model might lead to
probabilities which were actually \emph{correlated} with the property
of interest, requiring one to balance various competing effects.  This
is a very interesting possibility but one which requires a broader
discussion than we can make here.  Anyways, keeping this qualification
in mind, let us proceed.

Thus, we would like to compute $dN_{\rm vac}(M_{\rm susy})$ given the
observed values of $\Lambda \sim 0$ and $M_{EW} \sim 100 \, {\rm
GeV}$, with $M_{\rm susy}^4$ defined to be equal to the positive
definite contribution to the supergravity potential, {\it i.e.}
\begin{equation}\label{eq:susyscale}
M_{\rm susy}^4 \equiv \sum_a |F_a|^2 + \sum_A D_A^2
\end{equation}
where $F_i=e^{K/2}D_i W$.
Suppose this followed an approximate power law
distribution, $dN_{\rm vac}(M_{\rm susy})
\sim M_{\rm susy}^\alpha dM_{\rm susy}$, then for $\alpha<-1$ vacuum
statistics would favor low scale susy, while for $\alpha > -1$ it
would not.

For purposes of comparison, let us begin with the standard
prediction of field theoretic naturalness, implicit in the
motivation often given for low scale supersymmetry breaking based on
the hierarchy problem. This is
\begin{equation}\label{eq:FTnatural}
dN_{\rm vac}^{FT}(M_{\rm susy})
 \sim
\left( \frac{M_{EW}^2~M_{\rm Pl}^2}{M_{\rm susy}^4} \right) \left(
\frac{\Lambda}{M_{\rm susy}^4} \right) f(M_{susy}),
\end{equation}
where the first factor represents the electroweak scale tuning, and
the second one the cosmological constant tuning assuming a
supersymmetric vacuum has cosmological constant zero (as is the case
in rigid supersymmetric field theory). The factor $f(M)$ represents
the a priori distribution coming out of the underlying theory,
independent of these tuning requirements. If we grant that this is
set by strong gauge dynamics as in conventional field theory models
of supersymmetry breaking, a reasonable ansatz might be $f(M)=dM/M$,
analogous to Eq. (\ref{conifdistr}). This would lead to $\alpha=-9$ and
a clear statistical prediction.

On the other hand, this is leaving out all of the fluxes and
hidden sectors which were postulated in the Bousso-Polchinski model,
and are generic in actual string/M theory compactifications.
Since the expression \rfn{susyscale} is a sum
of squares, a simple model for their effect is that $M_{susy}$
receives many independent positive contributions,
about as many as there are fluxes or at least moduli,
leading to prior distributions of the form
\cite{Douglas:2004qg,Susskind:2004uv}
\begin{equation} \label{eq:highsusy}
 f(M_{\rm susy}) \sim d M_{\rm susy}^{4K} .
\end{equation}
This would lead to a large positive value for $\alpha$ and
overwhelmingly prefer high energy supersymmetry breaking.

Thus, depending on our microscopic picture, we arrive at very different
conclusions. However, both of these simple considerations have flaws.
The simplest problem with \rfn{FTnatural} is the factor
$\Lambda/M_{susy}^4$.  Instead, cosmological constant distributions
of flux vacua generally go as $\Lambda/M_F^4$, with $M_F = M_{KK}$
or $M_F=M_P$ or some other fundamental scale. In other words, the
tuning problem of the cosmological constant is not helped by
supersymmetry. Essentially the reason is that in supergravity, the
effective potential receives both positive and negative
contributions, with negative contributions $\sim -|W|^2$ persisting
for supersymmetric vacua.  In the case of flux vacua, these negative
contributions are distributed
roughly uniformly up to some fundamental scale $M_F^4$, independent
of the supersymmetry breaking scale, as we saw at the end of section
\ref{sec:stringfluxvac}. This leads to a tuning factor
$\Lambda/M_F^4$ instead of $\Lambda/M_{\rm susy}^4$ in
Eq. (\ref{eq:FTnatural}), resulting in $\alpha=-5$, still favoring low
scale supersymmetry, but rather less so.

There are many other implicit assumptions and even gaps in the
arguments for \rfn{FTnatural}, many already recognized in the
literature.  Among those which implicitly favor low scale breaking,
the expression \rfn{FTnatural} assumes a generic solution to the $\mu$
problem, as well as to the other known problems of supersymmetric
phenomenology such as FCNC's etc.  As an example on the other side,
there might exist some generic class of models in which the
supersymmetric contributions to $W$ are forced to be small, say
by postulating an R symmetry which is only broken along with
supersymmetry breaking, restoring the $\Lambda/M_{\rm susy}^4$ factor.

Let us however turn to the arguments leading to \rfn{highsusy}, which
if true would potentially outweigh all of these other considerations.
To examine this further, we need a microscopic model of supersymmetry
breaking.  In fact, one can expect a generic potential to contain
many metastable supersymmetry breaking minima,
not because of any ``mechanism,'' but simply because generic
functions have many minima.  Indeed this was shown to be generic for
IIb flux vacua in \cite{Denef:2004cf}, leading to the distribution
\begin{equation}\label{eq:branchone}
f(M_{\rm susy}) \sim d \left( \frac{M_{\rm susy}}{M_{F}}
\right)^{12}
\end{equation}
in the regime $M_{\rm susy} \ll M_F$. This would still favor high
superymmetry breaking scales in \rfn{FTnatural},
but much less so than \rfn{highsusy} -- the flaw in the argument for
the latter is that the different
contributions to $F^2=\sum_a |F_a|^2$ are \emph{not} independent,
but correlated by the critical point conditions $\partial_a V=0$.

Although the specific power $12$ may be surprising at first, it has
a simple explanation \cite{Dine:2005yq,Giudice:2006sn}. Let us
consider a generic flux vacuum with $M_{\rm susy} \ll M_{F}$. Since
one needs a goldstino for spontaneous susy breaking, at least one
chiral superfield must have a low mass; call it $\phi$. Generically,
the flux potential gives order $M_{F}$ masses to all the other
chiral superfields, so they can be ignored, and we can analyze the
constraints in terms of an effective superpotential reduced to
depend on the single field $\phi$,
\begin{equation}
W(\phi) = W_0 + a \phi + b \phi^2 + c \phi^3 + \ldots .
\end{equation}
The conditions for a metastable supersymmetric vacuum are then
$|a| = M_{\rm susy}^2$, $|b| = 2 |a|$ (this follows
from the equation $V'=0$), and finally $|c| \sim |a|$ (as explained
in \cite{Denef:2004cf} and many previous discussions, this is
necessary so that $V''>0$. This also requires a lower bound on the
curvature of the moduli space metric).

Furthermore, an analysis of flux superpotentials along the lines of
our previous discussion bears out the expectation that the parameters
$(a,b,c)$ are independent and uniformly distributed complex
parameters, up to the flux potential cutoff scale $M_{F}$.  To get low
scale breaking, all three complex parameters must be tuned to be small
in magnitude, leading directly to Eq. (\ref{eq:branchone}).

The flaw in applying a standard naturalness argument here is very
simple; one needs to tune several parameters in the microscopic theory
to accomplish a single tuning at the low scale.  Of course, if the
underlying dynamics correlated these parameters, one could recover
natural low scale breaking.  Thus, we have not replaced the paradigm
of naturalness, but rather sharpened and extended it.  But even
granting that theories with such dynamics exist, the question becomes
whether among the many possibilities contained in the string theory
landscape, the vacua realizing them are sufficiently numerous to dominate
the simpler ``fine tuning'' scenario.  After all,
the fine tuning we are trying to explain is only of order
$(100 \GeV/M_P)^2 \sim 10^{-34}$, and for all we know the fraction of
string theory vacua which do this by low scale supersymmetry breaking
is even smaller.  While it will probably be some time before we can
convincingly answer such questions, thinking about them has already
shed new light on many old problems.

Having a precise microscopic picture from string theory becomes
particularly important when one is evaluating the naturalness of
discrete choices, or trying to weigh the importance of competing
effects.  As an example, let us consider the possibility that some of
the problems we discussed (such as large $|W|$ leading to
$\Lambda/M_F^4$) could be solved by postulating a discrete R symmetry.
Indeed, almost all existing proposals for natural models of low scale
breaking, such as those discussed in
\cite{Dine:2006gm}, rely on this postulate.

Unfortunately for such proposals, there
is a simple argument that discrete R symmetry
is heavily disfavored in flux vacua
\cite{Dine:2005gz}.
First, a discrete symmetry which acts on Calabi-Yau moduli space,
will have fixed points corresponding to particularly symmetric
Calabi-Yau manifolds; at one of these, it acts as a discrete
symmetry of the Calabi-Yau.  Such a symmetry of the Calabi-Yau will
also act on the fluxes, trivially on some and non-trivially others.
To get a flux vacuum respecting the symmetry, one must turn on only
invariant fluxes.  Now, looking at examples, one finds that
typically an order one fraction of the fluxes transform
non-trivially; for definiteness let us say half of them. Thus,
applying Eq. (\ref{adformula}) and putting in some typical numbers for
definiteness, we might estimate
$$
\frac{N_{\rm vac~symmetric}}{N_{\rm vac~all}} \sim
 \frac{L^{K/2}}{L^{K}} \sim \frac{10^{100}}{10^{200}} .
$$
Thus, discrete symmetries of this type come with a huge penalty.
While one can imagine discrete symmetries with other origins for
which this argument might not apply, since $W$ receives flux
contributions, it clearly applies to the R symmetry desired in
branch (3), and probably leads to suppressions far outweighing the
potential gains.  (Fortunately, this does not apply to R parity).

This is in stark contrast to traditional naturalness considerations,
in which all symmetries are ``natural.''  And there are other
examples of string vacuum distributions which come out differently
from traditional expectations, or which are different among different
classes of string/M theory vacua, as discussed in
\cite{Douglas:2006es,Kumar:2006tn} and references there.

Other distributions are simply not predicted at all by traditional
naturalness arguments.  A primary example is the distribution of gauge
groups and charged matter content.  Suppose we were to
find evidence for a new strongly coupled gauge sector (perhaps
responsible for supersymmetry breaking, perhaps not), but had very
limited information about the matter spectrum, say a single resonance.
What should we expect for the gauge group?  Simple guesses might
be $SU(2)$ (the smallest nonabelian group), or $SU(4)$ (following
the pattern $1-2-3-\ldots$).  On the other hand, a huge amount of
theoretical work has been devoted to the proposition that $SU(N)$
gauge theory becomes simpler as $N$ becomes large; should we not give
this intuition equal weight?

While the considerations we just cited are unconvincing,
in fact string/M theory does predict a definite distribution of
gauge theories and matter contents, which has been explored in
numerous recent works, including
\cite{Blumenhagen:2004xx,Dijkstra:2004cc,Gmeiner:2005nh,Douglas:2006xy,Kumar:2005hf,Dienes:2006ut}
and many more cited in the reviews.  Besides bearing on the
susy breaking scale, such results could be useful in guiding searches
for exotic matter, in motivating other proposals for dark matter, and so on.

\section{Implications for the testability of string theory}
\label{sec:test}

As of 2007, it seems fair to say that while string theory remains by
far the best candidate we have for a complete theory of fundamental
physics, there is still no compelling empirical evidence for or
against the claim that the theory describes our universe.

Many different approaches have been proposed to look for such
evidence.  Perhaps the simplest, and certainly the best founded in
the history of our subject, is to look for ``exotic'' or
``signature'' physics which is easily modelled by string theory, and
not by other theories.  There are many such phenomena, such as
excitation states of the string (Regge recurrences).  Each is
associated with a new energy scale, for example the string scale
$M_s$ for excited string states.  As another example, in higher
dimensional theories, once one reaches energies of order the higher
dimensional Planck scale $M_P$, one can have black hole creation.
Presumably, Planck mass black holes decay very rapidly and this
process can be thought of as creating an unstable particle, however
it has been argued that the resulting distribution of decay products
will look very different from other interactions.  And, the higher
dimensional Planck scale can in principle be far lower than the
four-dimensional Planck scale, perhaps low enough to make this
observable (for a recent review, see \cite{Landsberg:2006mm}).

One can add the Kaluza-Klein scale $M_{KK}$ to the list; while
logically speaking the observation of extra dimensions is not a
direct test of string theory, clearly it would have an equally
profound significance for fundamental physics.  There are a number
of other ``exotic'' phenomena of which one can say the same thing,
such as the possibility of non-trivial four dimensional fixed point
field theories, and phenomena related to ``warping.''  Again, it is
not hard to propose scenarios in which this new physics will not be
seen until we reach a new energy scale $M_{EX}$.  Thus, the larger
problem is to decide whether any of these phenomena are relevant in
our universe.

The most direct approach to testing the theory is to find a way to
probe these energy scales.  At present, phenomenological constaints on
all of these new energy scales appear to be very weak, ranging from
just above current collider bounds, around a few $\TeV$, all the way
up to the GUT and four dimensional Planck scales $M \ge 10^{16} \GeV$.
In
most of this parameter space, the exotic phenomena may well be
inaccessible in terrestrial experiments and irrelevant in almost all
astrophysical processes.  Thus, while string theory can offer
experimentalists many exciting possibilities, there is little in the
way of guarantees, nor any clear way for such searches to \emph{falsify} the
theory.

Thus, a central question for string theorists is to better constrain
these scales theoretically.  Since parameters such as the size of the
extra dimensions are moduli, the considerations we have discussed in
our review are clearly very relevant for this, and there are already
many interesting suggestions.  For example, it appears that in
KKLT IIb flux vacua with many moduli, large extra dimensions are
disfavored.  On the other hand, there is an alternate regime in IIb
theory in which the structure of the effective potential requires
large extra dimensions \cite{Balasubramanian:2005zx}, and it appears
that other constructions such as IIa and M theory flux vacua may
statistically favor large extra dimensions
\cite{DeWolfe:2005uu,Acharya:2005ez}.
Thus the picture at present is not very clear; furthermore it seems
very likely {\it a priori} that considerations from early cosmology
bear on this particular question; but it is reasonable to expect
significant theoretical progress on this question.

Again, this progress is likely to lead to ``statistical predictions,''
in the sense that even if most vacua are shown to have some property
(say for sake of discussion, large extra dimensions), there will be
exceptions.  To make a perhaps evident comment on the value of this,
while in the hypothetical situation we discuss one would not be able to say
that ``ruling out'' large extra dimensions would falsify string
theory, one would at least know that one had drastically narrowed down
the possibilities, allowing one to go on to determine the most
promising next avenues for potential tests.  In this way, getting a
picture of the landscape is useful and perhaps even necessary
in guiding the search for conclusive tests.

Let us now turn to the question of testability if we do \emph{not} see
exotic physics. To be more precise, suppose that all observed physics
can be well described by some 4d effective field theory
coupled to gravity.  This is certainly true at present, and it might
well turn out so for physics at $1$ and even $10 \TeV$ as well.
While this would seem a frustrating possibility, one can certainly
still hope to make contact with string theory from such data.  After all,
we believe that string/M theory has a finite number of vacua
\cite{Acharya:2006zw},
and thus can lead to a finite number of 4d low energy theories;
could we imagine showing that the data is fit by \emph{none} of these
theories, thus falsifying the theory?

Approaches of this type include the following.
\begin{enumerate}
\item Find ``no-go'' arguments that certain phenomena, which can be
described by effective field theory, in fact cannot arise in string
theory.  For example, one can argue this for time variation of the
fine structure constant \cite{Banks:2001qc}.  One can also place bounds
on gauge couplings (admittedly, far from the observed values)
\cite{Arkani-Hamed:2006dz}.
\item Similarly, we could try to use the phenomena to
rule out competing theories, thereby getting circumstantial evidence
for string theory.  The basic example here is that, at present,
there is no other generally accepted theory of four dimensional
quantum gravity, and this is commonly taken as circumstantial
evidence for string theory.  Of course, one should not take this too
seriously until it can be proven that alternatives do not exist.  In
our opinion, the most promising alternative to study at present is
the idea that certain extended supergravity theories might provide
finite theories of gravity
\cite{Bern:2006kd,Green:2006yu}.
\item Perhaps physics at the few $\TeV$ scale of LHC and ILC
will turn out to show some remarkable simplicity which can easily be
reproduced by string theory compactification.  One idea is to focus
on properties of the particular
GUT theories obtained by heterotic string compactification.
More recently, it has been suggested that certain D-brane
``quiver gauge theories'' leading to Standard-like
models are preferred
\cite{Berenstein:2006pk,Verlinde:2005jr}.
\item Make statements about the ``likely'' distribution of predictions
among all vacua of string theory.  Then, to the extent that what we
see is ``likely,'' we again get circumstantial evidence for string
theory.
\end{enumerate}
Again, (1-3) suffer from the general problem that the
phenomena being discussed may not actually be properties of our
universe, while (4) suffers from the problem that we might just live
in an ``unlikely'' universe.  Thus, one would probably need to
combine information from all of these approaches to make progress.

A more optimistic version of (4) holds that a better understanding of
early cosmology and whatever mechanism populates the many vacua of the
theory, will lead to a strongly peaked probability distribution which
selects one or a handful of the candidate vacua in a way amenable to
calculation.  The search for gauge invariant and well-defined
inflationary measures has been a 20-year struggle; recent progress is
summarized in the short review \cite{Vilenkin:2006xv}.  We note that
even should one find a natural and computable candidate measure,
actually finding the preferred vacua may be a daunting problem.
It is not hard to imagine scenarios in which this is
impossible even in principle, because of fundamental limitations
coming from the theory of computational complexity
\cite{Denef:2006ad}.

It is hard at present to judge the prospects for any of these approaches.
From thinking about historical analogies,
a tentative conclusion for the
problem at hand is that while a great deal can be learned on the
theoretical side, perhaps eventually allowing us to propose a
definite test, ultimately convincing evidence for string theory will
probably have to come from observing some sort of exotic physics.
A natural place to look for this is early cosmology, as
the physics of inflation involves very high energies, with
$V \sim M_{GUT}^4$ in many models.  Several of the proposed models
of inflation in string theory have characteristic signatures which
(if sufficiently well measured) encode stringy
physics. These include networks of cosmic
D and F strings formed during the exit from brane inflation
\cite{Sarangi:2002yt,Copeland:2003bj,Dvali:2003zj} and
non-Gaussian signals in the CMB radiation which probe the specific
non-linearities of the DBI action
\cite{Alishahiha:2004eh,Babich:2004gb,Chen:2006nt}.

Space prohibits a detailed discussion of these and many other
interesting ideas.
We conclude by noting that while the present situation is not very
satisfactory, there is every reason to be optimistic.  In string/M theory,
we have a theoretical framework which on the one hand is grounded in
precise mathematics (so that many, even most theoretical suggestions can
be falsified on internal grounds), yet which on the other hand shows
significant promise of making contact with observable physics.
There are many well-motivated directions for improving the situation,
and good reasons to believe that substantial progress will be made
in the future.

\section*{Acknowledgements}

The work of M.R.D. was supported by DOE grant DE-FG02-96ER40959,
and the work of S.K. was supported in part by a David and Lucile Packard
Foundation Fellowship, the DOE under contract DE-AC02-76SF00515, and
the NSF under grant number 0244728.


\begin{thebibliography}{999}

\bibitem{Scherk:1974mc}
  J.~Scherk and J.~H.~Schwarz,
  Phys.\ Lett.\ B {\bf 52}, 347 (1974).

\bibitem{Schwarz:1982jn}
  J.~H.~Schwarz,
  Phys.\ Rept.\  {\bf 89}, 223 (1982).

\bibitem{Green:1984sg}
  M.~B.~Green and J.~H.~Schwarz,
  Phys.\ Lett.\ B {\bf 149}, 117 (1984).

\bibitem{Candelas:1985en}
  P.~Candelas, G.~T.~Horowitz, A.~Strominger and E.~Witten,
  Nucl.\ Phys.\ B {\bf 258}, 46 (1985).

\bibitem{Dine:1985he}
  M.~Dine and N.~Seiberg,
  Phys.\ Lett.\ B {\bf 162}, 299 (1985).

\bibitem{Sen:2001di}
  A.~Sen,
{\it Prepared for Les Houches Summer School: Session 76: Euro Summer
School on Unity of Fundamental Physics: Gravity, Gauge Theory and
Strings, Les Houches, France, 30 Jul - 31 Aug 2001}

\bibitem{Intriligator:1995au}
  K.~A.~Intriligator and N.~Seiberg,
  Nucl.\ Phys.\ Proc.\ Suppl.\  {\bf 45BC}, 1 (1996)
  [arXiv:hep-th/9509066].

\bibitem{Weinberg:1988cp}
  S.~Weinberg,
  Rev.\ Mod.\ Phys.\  {\bf 61}, 1 (1989).

\bibitem{Weinberg:2000yb}
  S.~Weinberg,
  arXiv:astro-ph/0005265.

\bibitem{Chanowitz:1988ae}
  M.~S.~Chanowitz,
  Ann.\ Rev.\ Nucl.\ Part.\ Sci.\  {\bf 38}, 323 (1988).

\bibitem{Adelberger:2003zx}
  E.~G.~Adelberger, B.~R.~Heckel and A.~E.~Nelson,
  Ann.\ Rev.\ Nucl.\ Part.\ Sci.\  {\bf 53}, 77 (2003)
  [arXiv:hep-ph/0307284].

\bibitem{Banks:1993en}
  T.~Banks, D.~B.~Kaplan and A.~E.~Nelson,
  Phys.\ Rev.\ D {\bf 49}, 779 (1994)
  [arXiv:hep-ph/9308292].

\bibitem{deCarlos:1993jw}
  B.~de Carlos, J.~A.~Casas, F.~Quevedo and E.~Roulet,
  Phys.\ Lett.\ B {\bf 318}, 447 (1993)
  [arXiv:hep-ph/9308325].

\bibitem{Coughlan:1983ci}
  G.~D.~Coughlan, W.~Fischler, E.~W.~Kolb, S.~Raby and G.~G.~Ross,
  Phys.\ Lett.\ B {\bf 131}, 59 (1983).

\bibitem{Peebles:2002gy}
  P.~J.~E.~Peebles and B.~Ratra,
  Rev.\ Mod.\ Phys.\  {\bf 75}, 559 (2003)
  [arXiv:astro-ph/0207347].

\bibitem{Banks:1984tw}
T.~Banks, Phys. Rev. Lett. {\bf 52}, 1461 (1984).

\bibitem{Linde:1984ir}
A.D. Linde, Rept. Prog. Phys. {\bf 47}, 925 (1984).

\bibitem{Weinberg:1987dv}
  S.~Weinberg,
  Phys.\ Rev.\ Lett.\  {\bf 59}, 2607 (1987).

\bibitem{Lerche:1986cx}
  W.~Lerche, D.~Lust and A.~N.~Schellekens,
  Nucl.\ Phys.\ B {\bf 287}, 477 (1987).

\bibitem{Bousso:2000xa}
  R.~Bousso and J.~Polchinski,
  JHEP {\bf 0006}, 006 (2000)
  [arXiv:hep-th/0004134].

\bibitem{Gukov:1999ya}
S.~Gukov, C.~Vafa and E.~Witten, Nucl. Phys. B {\bf 608}, 477 (2001)
[arXiv:hep-th/9906070].

\bibitem{Dasgupta:1999ss}
K.~Dasgupta, G.~Rajesh and S.~Sethi, JHEP {\bf 9908}, 023 (1999)
[arXiv:hep-th/9908088].

\bibitem{Mayr:2000hh}
P.~Mayr, Nucl. Phys. B {\bf 593}, 99 (2001) [arXiv:hep-th/0003198].

\bibitem{Curio:2000sc}
G.~Curio, A.~Klemm, D.~Lust and S.~Theisen, Nucl. Phys. B {\bf 609},
3 (2001) [arXiv:hep-th/0012213].

\bibitem{Giddings:2001yu}
S.B. Giddings, S. Kachru and J. Polchinski, Phys. Rev. D {\bf 66},
106006 (2002) [arXiv:hep-th/0105097].

\bibitem{Silverstein:2001xn}
E. Silverstein, arXiv:hep-th/0106209.

\bibitem{Kachru:2003aw}
  S.~Kachru, R.~Kallosh, A.~Linde and S.~P.~Trivedi,
  Phys.\ Rev.\ D {\bf 68}, 046005 (2003)
  [arXiv:hep-th/0301240].

\bibitem{Denef:2004dm}
F.~Denef, M.~R.~Douglas and B.~Florea, JHEP {\bf 0406}, 034 (2004)
[arXiv:hep-th/0404257].

\bibitem{Denef:2005mm}
F.~Denef, M.~R.~Douglas, B.~Florea, A.~Grassi and S.~Kachru,
arXiv:hep-th/0503124.

\bibitem{Lust:2006zg}
D.~Lust, S.~Reffert, E.~Scheidegger, W.~Schulgin and S.~Stieberger,
arXiv:hep-th/0609013.


\bibitem{Susskind:2003kw}
  L.~Susskind,
  arXiv:hep-th/0302219.

\bibitem{Douglas:2006es}
  M.~R.~Douglas and S.~Kachru,
  arXiv:hep-th/0610102.

\bibitem{Blumenhagen:2006ci}
R.~Blumenhagen, B.~Kors, D.~Lust and S.~Stieberger,
arXiv:hep-th/0610327.


\bibitem{Grana:2005jc}
  M.~Grana,
  Phys.\ Rept.\  {\bf 423}, 91 (2006)
  [arXiv:hep-th/0509003].

\bibitem{Silverstein:2004id}
  E.~Silverstein,
  arXiv:hep-th/0405068.

\bibitem{Frey:2003tf}
  A.~R.~Frey,
  arXiv:hep-th/0308156.

\bibitem{Freund:1980xh}
  P.~G.~O.~Freund and M.~A.~Rubin,
  Phys.\ Lett.\ B {\bf 97}, 233 (1980).

\bibitem{McGreevy:2006hk}
  J.~McGreevy, E.~Silverstein and D.~Starr,
  arXiv:hep-th/0612121.

\bibitem{Saltman:2004jh}
  A.~Saltman and E.~Silverstein,
  JHEP {\bf 0601}, 139 (2006)
  [arXiv:hep-th/0411271].

\bibitem{DeWolfe:2005uu}
  O.~DeWolfe, A.~Giryavets, S.~Kachru and W.~Taylor,
  JHEP {\bf 0507}, 066 (2005)
  [arXiv:hep-th/0505160].

\bibitem{Villadoro:2005cu}
  G.~Villadoro and F.~Zwirner,
  JHEP {\bf 0506}, 047 (2005)
  [arXiv:hep-th/0503169].

\bibitem{Camara:2005dc}
P.G. Camara, A. Font and L.E. Ibanez, JHEP {\bf 0509}, 013 (2005)
[arXiv:hep-th/0506066].


\bibitem{Dixon:1985jw}
  L.~J.~Dixon, J.~A.~Harvey, C.~Vafa and E.~Witten,
  Nucl.\ Phys.\ B {\bf 261}, 678 (1985).

\bibitem{Banks:2006hg}
  T.~Banks and K.~van den Broek,
  arXiv:hep-th/0611185.

\bibitem{Berkooz:1996km}
  M.~Berkooz, M.~R.~Douglas and R.~G.~Leigh,
  Nucl.\ Phys.\ B {\bf 480}, 265 (1996)
  [arXiv:hep-th/9606139].

\bibitem{Blumenhagen:2005mu}
  R.~Blumenhagen, M.~Cvetic, P.~Langacker and G.~Shiu,
  Ann.\ Rev.\ Nucl.\ Part.\ Sci.\  {\bf 55}, 71 (2005)
  [arXiv:hep-th/0502005].

\bibitem{Intriligator:2006dd}
  K.~Intriligator, N.~Seiberg and D.~Shih,
  JHEP {\bf 0604}, 021 (2006)
  [arXiv:hep-th/0602239].

\bibitem{Ooguri:2006bg}
  H.~Ooguri and Y.~Ookouchi,
  Phys.\ Lett.\ B {\bf 641}, 323 (2006)
  [arXiv:hep-th/0607183].

\bibitem{Kumar:2006tn}
  J.~Kumar,
  Int.\ J.\ Mod.\ Phys.\ A {\bf 21}, 3441 (2006)
  [arXiv:hep-th/0601053].

\bibitem{Douglas:2003um}
  M.~R.~Douglas,
  JHEP {\bf 0305}, 046 (2003)
  [arXiv:hep-th/0303194].

\bibitem{Ashok:2003gk}
  S.~Ashok and M.~R.~Douglas,
  JHEP {\bf 0401}, 060 (2004)
  [arXiv:hep-th/0307049].

\bibitem{Denef:2004ze}
  F.~Denef and M.~R.~Douglas,
  JHEP {\bf 0405}, 072 (2004)
  [arXiv:hep-th/0404116].

\bibitem{Denef:2004cf}
  F.~Denef and M.~R.~Douglas,
  JHEP {\bf 0503}, 061 (2005)
  [arXiv:hep-th/0411183].

\bibitem{Klebanov:2000hb}
  I.~R.~Klebanov and M.~J.~Strassler,
  JHEP {\bf 0008}, 052 (2000)
  [arXiv:hep-th/0007191].

\bibitem{Randall:1999ee}
  L.~Randall and R.~Sundrum,
  Phys.\ Rev.\ Lett.\  {\bf 83}, 3370 (1999)
  [arXiv:hep-ph/9905221].

\bibitem{Kachru:2002gs}
  S.~Kachru, J.~Pearson and H.~L.~Verlinde,
  JHEP {\bf 0206}, 021 (2002)
  [arXiv:hep-th/0112197].

\bibitem{Kachru:2003sx}
  S.~Kachru, R.~Kallosh, A.~Linde, J.~M.~Maldacena, L.~McAllister and S.~P.~Trivedi,
  JCAP {\bf 0310}, 013 (2003)
  [arXiv:hep-th/0308055].

\bibitem{Hebecker:2006bn}
  A.~Hebecker and J.~March-Russell,
  arXiv:hep-th/0607120.

\bibitem{Douglas:2004qg}
  M.~R.~Douglas,
  arXiv:hep-th/0405279.

\bibitem{Susskind:2004uv}
  L.~Susskind,
  arXiv:hep-th/0405189.

\bibitem{Dine:2005yq}
  M.~Dine, D.~O'Neil and Z.~Sun,
  JHEP {\bf 0507}, 014 (2005)
  [arXiv:hep-th/0501214].

\bibitem{Giudice:2006sn}
  G.~F.~Giudice and R.~Rattazzi,
  Nucl.\ Phys.\ B {\bf 757}, 19 (2006)
  [arXiv:hep-ph/0606105].

\bibitem{Dine:2006gm}
  M.~Dine, J.~L.~Feng and E.~Silverstein,
  Phys.\ Rev.\ D {\bf 74}, 095012 (2006)
  [arXiv:hep-th/0608159].

\bibitem{Dine:2005gz}
  M.~Dine and Z.~Sun,
  JHEP {\bf 0601}, 129 (2006)
  [arXiv:hep-th/0506246].

\bibitem{Blumenhagen:2004xx}
  R.~Blumenhagen, F.~Gmeiner, G.~Honecker, D.~Lust and T.~Weigand,
  Nucl.\ Phys.\ B {\bf 713}, 83 (2005)
  [arXiv:hep-th/0411173].

\bibitem{Dijkstra:2004cc}
  T.~P.~T.~Dijkstra, L.~R.~Huiszoon and A.~N.~Schellekens,
  Nucl.\ Phys.\ B {\bf 710}, 3 (2005)
  [arXiv:hep-th/0411129].

\bibitem{Gmeiner:2005nh}
  F.~Gmeiner,
  Fortsch.\ Phys.\  {\bf 54}, 391 (2006)
  [arXiv:hep-th/0512190].

\bibitem{Douglas:2006xy}
  M.~R.~Douglas and W.~Taylor,
  arXiv:hep-th/0606109.

\bibitem{Kumar:2005hf}
  J.~Kumar and J.~D.~Wells,
  JHEP {\bf 0509}, 067 (2005)
  [arXiv:hep-th/0506252].

\bibitem{Dienes:2006ut}
  K.~R.~Dienes,
  Phys.\ Rev.\ D {\bf 73}, 106010 (2006)
  [arXiv:hep-th/0602286].

\bibitem{Landsberg:2006mm}
G. Landsberg, J. Phys. G {\bf 32}, R337 (2006)
[arXiv:hep-ph/0607297].


\bibitem{Balasubramanian:2005zx}
  V.~Balasubramanian, P.~Berglund, J.~P.~Conlon and F.~Quevedo,
  JHEP {\bf 0503}, 007 (2005)
  [arXiv:hep-th/0502058].

\bibitem{Acharya:2005ez}
  B.~S.~Acharya, F.~Denef and R.~Valandro,
  JHEP {\bf 0506}, 056 (2005)
  [arXiv:hep-th/0502060].

\bibitem{Acharya:2006zw}
  B.~S.~Acharya and M.~R.~Douglas,
  arXiv:hep-th/0606212.

\bibitem{Banks:2001qc}
  T.~Banks, M.~Dine and M.~R.~Douglas,
  Phys.\ Rev.\ Lett.\  {\bf 88}, 131301 (2002)
  [arXiv:hep-ph/0112059].

\bibitem{Arkani-Hamed:2006dz}
  N.~Arkani-Hamed, L.~Motl, A.~Nicolis and C.~Vafa,
  arXiv:hep-th/0601001.

\bibitem{Bern:2006kd}
  Z.~Bern, L.~J.~Dixon and R.~Roiban,
  arXiv:hep-th/0611086.

\bibitem{Green:2006yu}
  M.~B.~Green, J.~G.~Russo and P.~Vanhove,
  arXiv:hep-th/0611273.

\bibitem{Vilenkin:2006xv}
  A.~Vilenkin,
  arXiv:hep-th/0609193.

\bibitem{Denef:2006ad}
  F.~Denef and M.~R.~Douglas,
  arXiv:hep-th/0602072.

\bibitem{Berenstein:2006pk}
  D.~Berenstein and S.~Pinansky,
  arXiv:hep-th/0610104.

\bibitem{Verlinde:2005jr}
  H.~Verlinde and M.~Wijnholt,
  arXiv:hep-th/0508089.

\bibitem{Sarangi:2002yt}
  S.~Sarangi and S.~H.~H.~Tye,
  Phys.\ Lett.\ B {\bf 536}, 185 (2002)
  [arXiv:hep-th/0204074].

\bibitem{Copeland:2003bj}
  E.~J.~Copeland, R.~C.~Myers and J.~Polchinski,
  JHEP {\bf 0406}, 013 (2004)
  [arXiv:hep-th/0312067].

\bibitem{Dvali:2003zj}
  G.~Dvali and A.~Vilenkin,
  JCAP {\bf 0403}, 010 (2004)
  [arXiv:hep-th/0312007].

\bibitem{Alishahiha:2004eh}
  M.~Alishahiha, E.~Silverstein and D.~Tong,
  Phys.\ Rev.\ D {\bf 70}, 123505 (2004)
  [arXiv:hep-th/0404084].

\bibitem{Babich:2004gb}
  D.~Babich, P.~Creminelli and M.~Zaldarriaga,
  JCAP {\bf 0408}, 009 (2004)
  [arXiv:astro-ph/0405356].

\bibitem{Chen:2006nt}
  X.~Chen, M.~x.~Huang, S.~Kachru and G.~Shiu,
  arXiv:hep-th/0605045.


\end{thebibliography}
\end{document}